\documentclass[3p, times]{elsarticle}

\usepackage{amsmath}
\usepackage{amssymb}

\usepackage{makecell}

\usepackage[utf8]{inputenc}
\usepackage{xcolor}
\usepackage{graphicx}
\usepackage{bm}
\usepackage{ulem}

\newcommand{\taunl}{\tau_{\rm NL}}

\newcommand{\lcdm}{\Lambda{\rm CDM}}
\newcommand{\slcdm}{{\rm Super-}\lcdm}

\newcommand{\kmsMpc}{{\,\rm km/s/Mpc}}
\newcommand{\omk}{\Omega_k}

\journal{Physics of the Dark Universe}

\begin{document}
\title{The Hubble tension in the non-flat ${\rm Super-}\Lambda$CDM model}
\author{Saroj Adhikari}
\address{Department of Physics and Astronomy, State University of New York at Plattsburgh, Plattsburgh, NY, USA}
\ead{saroj.adhikari@plattsburgh.edu}

\begin{abstract}
    We investigate the Hubble tension in the non-flat ${\rm Super-}\Lambda$CDM model. The non-flat ${\rm Super-}\Lambda$CDM model extends the ${\rm Super-}\Lambda$CDM model by including the spatial curvature  as a free parameter. The ${\rm Super-}\Lambda$CDM model extends the standard $\Lambda$CDM model of cosmology through additional parameters accounting for the possible effect of a trispectrum in the primordial fluctuations. In the cosmic microwave background data, this effect can be parameterized using parameters that change the observed angular power spectrum from the theoretical power spectrum due to a trispectrum that couples long and short wavelength modes. In this work, we perform Markov Chain Monte Carlo (MCMC) data analysis on the recent {\it Planck} 2018 temperature and polarization fluctuations data and the local Hubble constant measurements from supernovae data assuming a non-flat ${\rm Super-}\Lambda$CDM model. We find that there is a preference for non-zero values of the spatial curvature parameter $\Omega_k$ and the ${\rm Super-}\Lambda$CDM parameter $A_0$ at a level of $\Delta \chi^2$ improvement of approximately 23.
\end{abstract}
\maketitle

\section{Introduction}
The {\it Planck} 2018 temperature and polarization fluctuations data currently provide very tight constraints on the standard $\lcdm$ model of cosmology \cite{Planck:2018vyg}. One of the important cosmological parameters that can be inferred from the Planck cosmic microwave background (CMB) fluctuations data is the Hubble constant \cite{Freedman:2010xv}, $H_0$, the current expansion rate of the universe. The {\it Planck} 2018 temperature and polarization data, assuming the standard $\lcdm$ model, provide constraints ($H_0 = 67.27\pm 0.60\ \kmsMpc$) that are tighter than the direct measurements of the Hubble constant using Type Ia supernovae from the SH0ES collaboration ($H_0 = 73.2\pm 1.3\ \kmsMpc$) \cite{Riess:2020fzl}. These measurements of the Hubble constant, $H_0$, from the early and late universe disagree at more than $4\sigma$ \cite{Verde:2019ivm}; see also \cite{Freedman:2021ahq}, however. A more recent work from the SH0ES team \cite{Riess:2021jrx} presents a tighter measurement of the Hubble constant at $H_0 = 73.04\pm 1.04\ \kmsMpc$ which brings the disagreement with the {\it Planck} CMB prediction to $5\sigma$. This disagreement between the measurement of Hubble constant from Type Ia supernovae and that derived from CMB is a very active area of investigation in cosmology. It is possible that the resolution of the discrepancy can be due to currently unaccounted for systematic effects in one or more of the data sets that are used in the analyses. It is also possible that the resolution of the discrepancy comes through an extension to the cosmological model. See recent review articles \cite{Mortsell:2018mfj, DiValentino:2021izs, Schoneberg:2021qvd} for discussions of several proposed solutions to the Hubble tension problem.

One potential solution to the Hubble tension was presented in \cite{Adhikari:2019fvb}; it extends the $\lcdm$ model by allowing the primordial fluctuations to deviate from the assumption of Gaussianity. The extended model is called the $\slcdm$ model and it requires the primordial fluctuations that seed the CMB fluctuations have a deviation from Gaussianity. In \cite{Adhikari:2019fvb}, using {\it Planck} 2015 temperature fluctuations data, it was shown that the $\slcdm$ model can alleviate the Hubble tension; the model reduced the tension ($2.5\sigma$), bud did not completely resolve it. Most of the potential solutions to the Hubble tension problem presented so far similarly fit this description: they are able to reduce the level of tension but unable to solve the tension completely. 

It is therefore interesting to consider if we can add another parameter to the $\lcdm$ model on top of the additional parameter(s) of the $\slcdm$ model, and whether such an analysis can further alleviate the Hubble tension. We do such an analysis in this work by adding the spatial curvature as a free parameter in the $\slcdm$ model. We investigate parameter constraints on the spatial curvature $\omk$ and the $\slcdm$ parameters $A_0, \epsilon$ using the final {\it Planck} satellite CMB temperature+polarization data in combination with the Riess et. al 2020 Hubble constant measurement \cite{Riess:2020fzl}. In the non-flat $\lcdm$ model, the expansion history of the universe depends on the curvature density parameter today ($\omk$) in addition to matter ($\Omega_m$), radiation ($\Omega_r$) and dark energy ($\Omega_\Lambda$) densities:
\begin{align}
	\left( \frac{H}{H_0} \right)^2 &= \omk (1+z)^2 + \Omega_m (1+z)^3 + \Omega_r(1+z)^4 + \Omega_\Lambda
\end{align}

The $\slcdm$ model considers the effect of a primordial trispectrum on the observed angular power spectrum. In the presence of a trispectrum, it is possible that long-wavelength fluctuations modulate the small scale modes and therefore the observed power spectrum. The effect can be modeled by modifying the theoretical angular power spectrum ($C_\ell$) in the following manner \cite{Adhikari:2019fvb}: 
\begin{align}{C}_\ell \rightarrow C_\ell + A_0 C_\ell(n_s+\epsilon),
\label{eq:Clobs}
\end{align}
where $A_0$ and $\epsilon$ are additional parameters of the $\slcdm$ model, and $C_\ell(n_s+\epsilon)$ are the angular power spectra evaluated by changing the spectral index from $n_s$ to $n_s+\epsilon$ with the other cosmological parameters fixed. It can be shown \cite{Adhikari:2019fvb} that the effect of an additional modulation term such as $A_0 C_\ell(n_s+\epsilon)$ on a pseudo-$C_\ell$ power spectrum estimator is equivalent to a non-Gaussian term in the covariance matrix due to a primordial trispectrum of the form \cite{Assassi:2012zq}: 
\begin{align}
	T(\bf{k_1}, {\bf k_2}, {\bf k_3}, {\bf k_4}) &= 4 \taunl \left( \frac{K}{\sqrt{k_1 k_3}}\right)^{-2\epsilon} P_\Phi(k_1) P_\Phi(k_3) P_\Phi(K),
	\label{eq:trispectrum}
\end{align}
where $P_\Phi(k) = (2\pi^2 A_\Phi/k^3)(k/k_0)^{n_s-1}$ is the power spectrum of potential fluctuations, $k_i$s are the four wavenumbers of the trispectrum in momentum space such that $\bf{k}_1+\bf{k}_2+\bf{k}_3+\bf{k}_4=0$ and $K = |\bf{k}_1-\bf{k}_2|=|\bf{k}_3-\bf{k}_4|$.
A negative $A_0$ parameter (as is preferred by the data analysis presented later) means that the actual value of the amplitude of fluctuations is larger than what is inferred from CMB power spectra assuming $\lcdm$. However, note that in the $\slcdm$ model the exact value of $A_0$ cannot be calculated given a trispectrum. Only the variance of $A_0$ i.e. $\langle A_0^2 \rangle$ can be calculated. As such, it is not easy to directly translate the constraints obtained on $A_0$ and $\epsilon$ parameters in this work to the trispectrum amplitude $\taunl$ of the primordial trispectrum. The $\epsilon$ parameter in the primrodial trispectrum Eq. (\ref{eq:trispectrum}) is related to the mass of the additional scalar field in the quasi-single field model that generates the trispectrum \cite{Chen:2009we, Chen:2009zp}. From Eq.(\ref{eq:Clobs}), we can see that $\epsilon$ has degeneracy with the spectral index $n_s$.

The modified angular power spectra in Eq.(\ref{eq:Clobs}) are used to fit with the experimental angular power spectra. In the MCMC sampling, the cosmological parameters, the calibration parameters and the $\slcdm$ parameters $(A_0, \epsilon)$ are all sampled together. To calculate the theoretical power spectra, $C_\ell$, we make use of {\tt camb} \cite{Lewis:1999bs, Howlett:2012mh}. To sample the posterior parameter distribution, we use the {\tt cobaya} \cite{Torrado:2020dgo} package and its implementation of the Metropolis sampler \cite{Lewis:2002ah, Lewis:2013hha, Neal:2005}. With the assumption that the long-wavelength modes that are responsible for the exact realization of both the temperature and polarization fluctuations of the Planck satellite data are mostly the same, we use a single $A_0$ parameter for all $TT, TE, EE$ power spectra. The amount of overlap of these long-wavelength modes with the CMB lensing power spectra $\phi\phi$ needs to be studied carefully, and as such, we will omit using the CMB lensing power spectra in this work.

There have been a few analyses so far that have added the spatial curvature as an additional parameter to the $\lcdm$ cosmological model and its extensions in the context of current discussions of cosmic parameter tensions \cite{Yang:2021hxg, DiValentino:2020hov, Vagnozzi:2020rcz, Shimon:2020dvb, Bose:2020cjb, KiDS:2020ghu,Shimon:2020dvb,Handley:2019tkm}, but in most of the cosmological analyses the universe is assumed to be flat. It is known that the {\it Planck}-only data favors a negative value of $\Omega_k$ at roughly $3\sigma$ \cite{DiValentino:2019qzk}; this preference is not robust to addition of CMB lensing and BAO data \cite{Planck:2018vyg}.

\section{Data Used}
\label{sec:data}
The main (most constraining) data we use are the Planck 2018 temperature and polarization fluctuations data. There are three separate likelihood calculations for these data, which are listed below. The descriptions of the {\it Planck} 2018 data and likelihoods are provided in \cite{Aghanim:2019ame}. Next, we use the Hubble constant measurement from the SH0ES collaboration \cite{Riess:2020fzl}, which disagrees with the derived $H_0$ from the Planck data at a significance greater than $4\sigma$ \cite{Verde:2019ivm}. When we allow $\omk$ to vary, the CMB-only (Planck) data does not constrain $\Omega_k$ very well. Therefore, our main result compares the MCMC results for two models ($\lcdm$ and non-flat $\slcdm$) when all of the following data sources are used:

\begin{itemize}
	\item Planck 2018 low temperature multipoles - {\tt planck\_2018\_lowl.TT}
	\item Planck 2018 low polarization multipoles - {\tt planck\_2018\_lowl.EE}
	\item Planck 2018 high temperature and polarization multipoles - {\tt planck\_2018\_highl\_plik.TTTEEE}
	\item SH0ES Hubble constant measurement - {\tt H0.riess2020} \cite{Riess:2020fzl}
	\item Pantheon Supernova data - {\tt SN.pantheon} \cite{Scolnic:2017caz}
\end{itemize}

The Pantheon sample consists of 1048 type Ia supernovae distance measurements in the redshift range $0.01<z<2.3$ \cite{Scolnic:2017caz}. We can add the Pantheon supernova data to our analysis without making any change to the available Pantheon likelihood. This is because the model prediction of distance measurements only depend on background parameters: $\Omega_m$, $\Omega_\Lambda$, $\Omega_k$, and $H_0$, and not on the additional $\slcdm$ parameters $A_0$ and $\epsilon$. We omit some cosmological datasets such as the Baryon Acoustic Oscillation (BAO) and CMB lensing for which detailed understanding of the theoretical prediction of the $\slcdm$ model is lacking. It will be interesting to check whether the result we obtain in this work holds when BAO and CMB lensing predictions for $\slcdm$ model are worked out and the relevant data are included.

In addition to the comparison of $\lcdm$ and non-flat $\slcdm$, we will also briefly discuss MCMC results for the non-flat $\lcdm$ model, in which $\omk$ is allowed to be a free parameter in addition to the six $\lcdm$ cosmological parameters.

\begin{table}
	\centering
	\begin{tabular}{| c | c | c | c |}
		\hline
		Parameter & Best-fit & Constraint (68\%) & 99.7\% $(3\sigma)$ Range \\
		\hline
		$A_0$  & $-0.168$ & $-0.169\pm 0.045$ & [-0.299, -0.040] \\
		$\omk$ & $0.00666$  & $0.0075_{-0.0023}^{+0.0025}$ & [-0.0003, 0.0140] \\
		$H_0$  & $72.08$  & $72.30 \pm 1.21$ & [68.72, 75.85] \\
		$\Omega_m$ & $0.2718$ & $0.2713\pm 0.0095$ & [0.2454, 0.3009] \\
		$100\theta_{\rm MC}$ & $1.04116$ & $1.04109\pm 0.00033$ & [1.04008, 1.04206] \\
		$\Omega_b h^2$ & $0.02266$ & $0.02260\pm 0.00018$ & [0.02210, 0.02310] \\
		$\log(10^{10}A_s)$ & $3.220$ & $3.221\pm0.051$ & [3.078, 3.375]\\
		$n_s$ & $0.9487$ & $0.946_{-0.014}^{+0.017}$ & [0.9043, 0.9787]\\
		$\epsilon$ & $-0.117$ & $-0.120_{-0.059}^{+0.088}$ & [-0.334, 0]\\
		\hline\hline
	\end{tabular}
	\caption{1D Marginalized constraint on some parameters of interest in the non-flat $\slcdm$ model fits performed using the ${\it Planck+H_0+Pantheon}$ data. The second column shows mean values and $1\sigma$ constraint whereas the third column shows the $3\sigma$ range of the parameter posteriors.}
	\label{table:parameters}
\end{table}

\section{Results}
\label{sec:results}
In Table \ref{table:parameters}, we list 1D marginalized posterior constraints for several parameters of the $\slcdm + \omk$ model. The second column gives the best-fit parameter values whereas the third column gives the posterior mean values with the corresponding one standard deviation constraint on a parameter. The table, in the third column, lists the three standard deviation range of posterior values for each of the parameters. 

\subsection{Constraint on $\lcdm$ parameters}
\begin{figure*}
	\centering
	\includegraphics[width=0.8\textwidth]{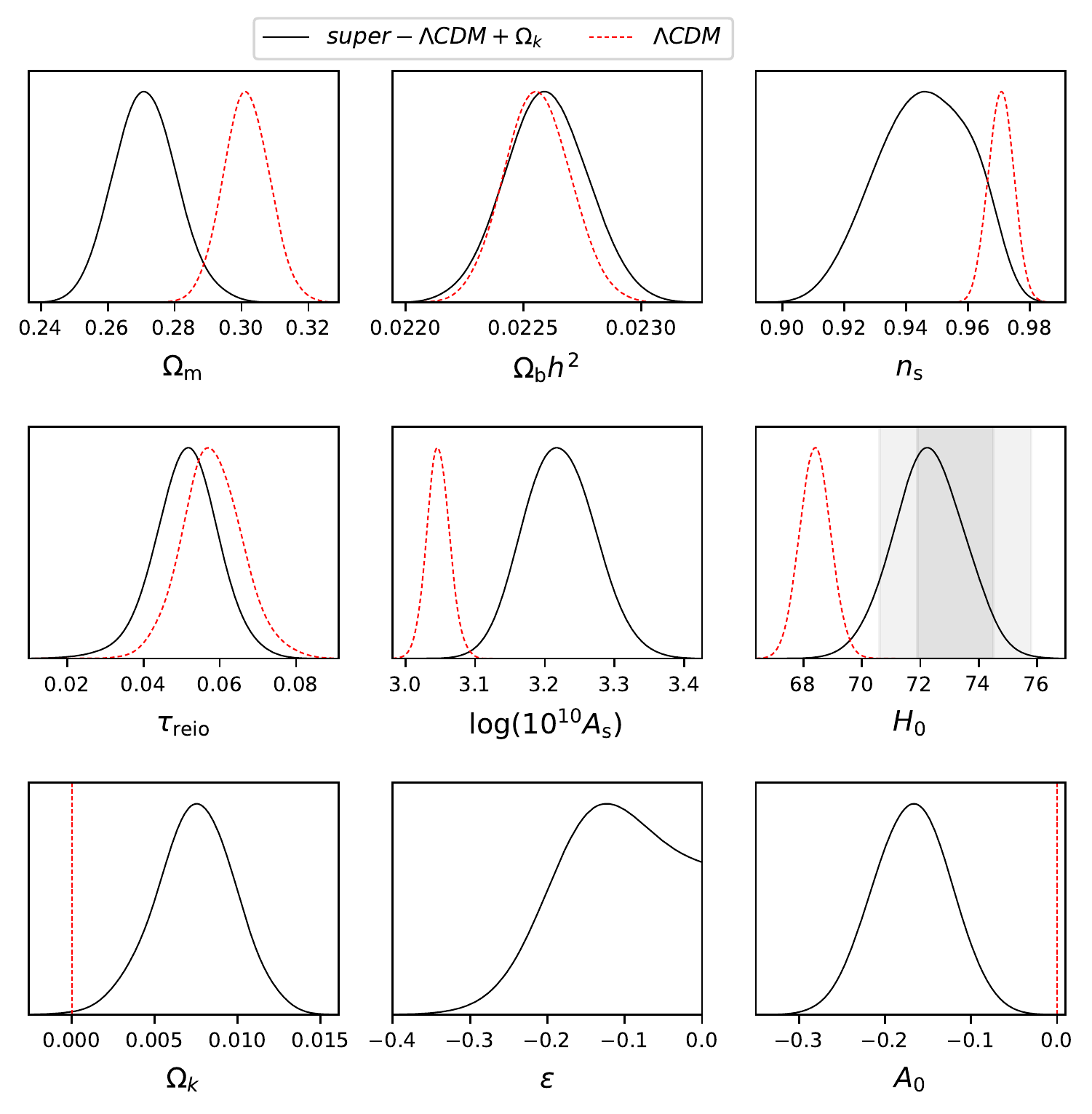}
	\caption{Top and middle panels: 1D marginalized constraint on six $\lcdm$ parameters in two models: standard $\lcdm$ model (red dashed) and the extended $\slcdm$ model (solid black). Relatively large shifts in parameter constraint values are between the two models are found for the matter density, the Hubble constant and the amplitude of fluctuations. The values of $H_0$ posterior in the $\slcdm+\omk$ model (solid black) are larger compared to the $\lcdm$ derived value of $H_0$ (red dashed), but are consistent with the SHOES $H_0$ measurement shown as gray bands (1 and 2 $\sigma$) in the middle panel. Bottom panel: 1D marginalized constraint on three new parameters in the non-flat $\slcdm$ model. In all cases, the data set used in our main data combination: {\it Planck + SH0ES + Pantheon}. The constraint on $\epsilon$ is weak as it has a strong degeneracy with $n_s$. On the other hand, the data combination generally prefers values of $\Omega_k$ and $A_0$ away from the $\lcdm$ null value of zero. See also Table \ref{table:parameters}.}
	\label{fig:parameters1D}
\end{figure*}

In Figure \ref{fig:parameters1D}, we show 1D marginalized posterior distribution for six $\lcdm$ parameters for our MCMC results from both $\lcdm$ (red dashed) and $\slcdm+\omk$ (black solid) models. Large parameter shifts occur for the Hubble constant ($H_0$), which is desired as our study of the non-flat $\slcdm$ model is motivated to solve the Hubble tension. When the spatial curvature is made a free parameter, the {\it Planck} and Pantheon data do not simultaneously constrain $\Omega_m, \Omega_k$ and $H_0$ parameters; the situation improves when the SH0ES Hubble constant measurement is included and the combined data set prefers a higher value of the Hubble constant in the non-flat $\slcdm$ model: $H_0 = 72.30\pm 1.21\ (68\ \%)$. 

\begin{figure*}
	\centering
	\includegraphics[width=0.85\textwidth]{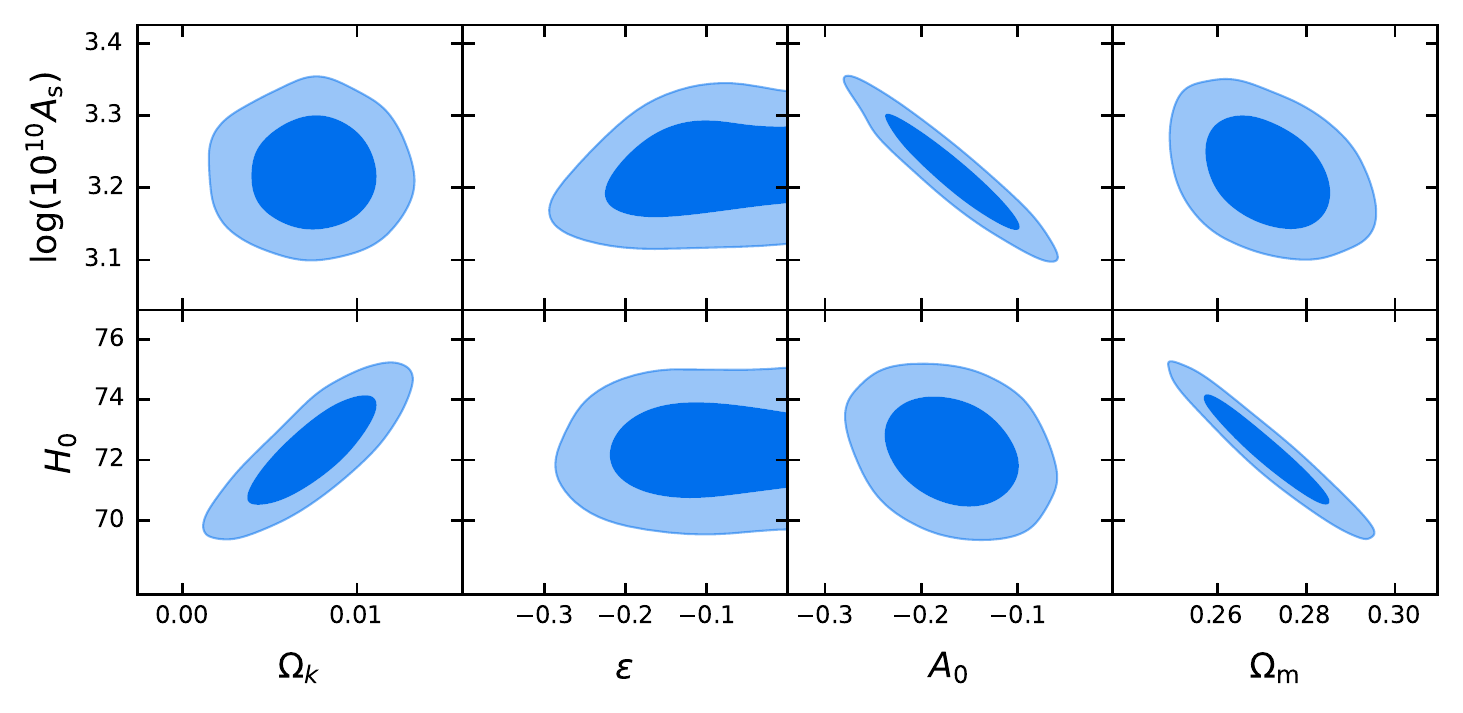}
	\caption{2D posterior probability distributions for the parameters $(A_0, \epsilon, \omk)$ sampled in our non-flat $\slcdm$ model fits, with the Hubble constant parameter, $H_0$. The data sets used are the {\it Planck} 2018 temperature and polarization CMB data, the Pantheon supernovae data and the Riess et. al Hubble constant measurement of \cite{Riess:2019cxk}. We also show the 2D posterior probability distributions of these parameters with the parameter for amplitude of fluctuations $\log(10^{10}A_s)$ in the top panel.}
	\label{fig:rectangle}
\end{figure*}

With a shift in the expansion rate to a higher value, there is a corresponding shift in the matter density to a lower value in the non-flat $\slcdm$ model: $\Omega_m = 0.2713\pm 0.0095\ (68\ \%)$. The strong degeneracy between the Hubble constant $H_0$ and the matter density $\Omega_m$ can be seen in the 2D posterior probability density plot of Figure \ref{fig:rectangle}.

There is also a preference in the non-flat $\slcdm$ model for the amplitude of fluctuations $A_s$ and therefore $\log(10^{10} A_s)$ to a higher value than in the $\lcdm$ model; see Figure \ref{fig:parameters1D}. This behavior is consistent with the previous work in \cite{Adhikari:2019fvb} and is due to the preference for a negative value of the $\slcdm$ parameter $A_0$; the strong degeneracy between $A_0$ and $\log(10^{10} A_s)$ can be seen in Figure \ref{fig:rectangle}.

\subsection{1D Marginalized constraint on $A_0, \epsilon$ and $\omk$}
The constraint on $A_0$, marginalized over all other parameters, is $A_0 = -0.169{\pm 0.045}\ (68\%)$. Compared to the constraint in \cite{Adhikari:2019fvb} where the $\omk$ parameter was fixed to zero, the marginalized constraint on $A_0$ in this work is tighter --- preferring non-zero value of $A_0$ at approximately $3\sigma$ even when marginalized over all other parameters. The 1D marginalized posterior distribution of the three new parameters in the extended model ($A_0$, $\epsilon$ and $\omk$) is shown in Figure \ref{fig:parameters1D} (bottom panel). Note that the data we used do not constrain the scale parameter $\epsilon$ of the $\slcdm$ model very well. The 1D marginalized constraint on $\omk$ is $\omk = 0.0075_{-0.0023}^{+0.0025}$, with a preference for a non-zero value of spatial curvature at slightly less than $3\sigma$ once we account for the  non-Gaussian nature of the 1D marginalized distribution. The posteriors for $\omk$ and $A_0$ are not significantly correlated. Next, we discuss how their joint posterior probability distribution significantly excludes the $\lcdm$ values of $(A_0=0, \omk=0)$. 

\subsection{2D Posterior distribution on $(A_0, \omk)$}

\begin{figure}[h]
	\centering
    \includegraphics[width=0.5\textwidth]{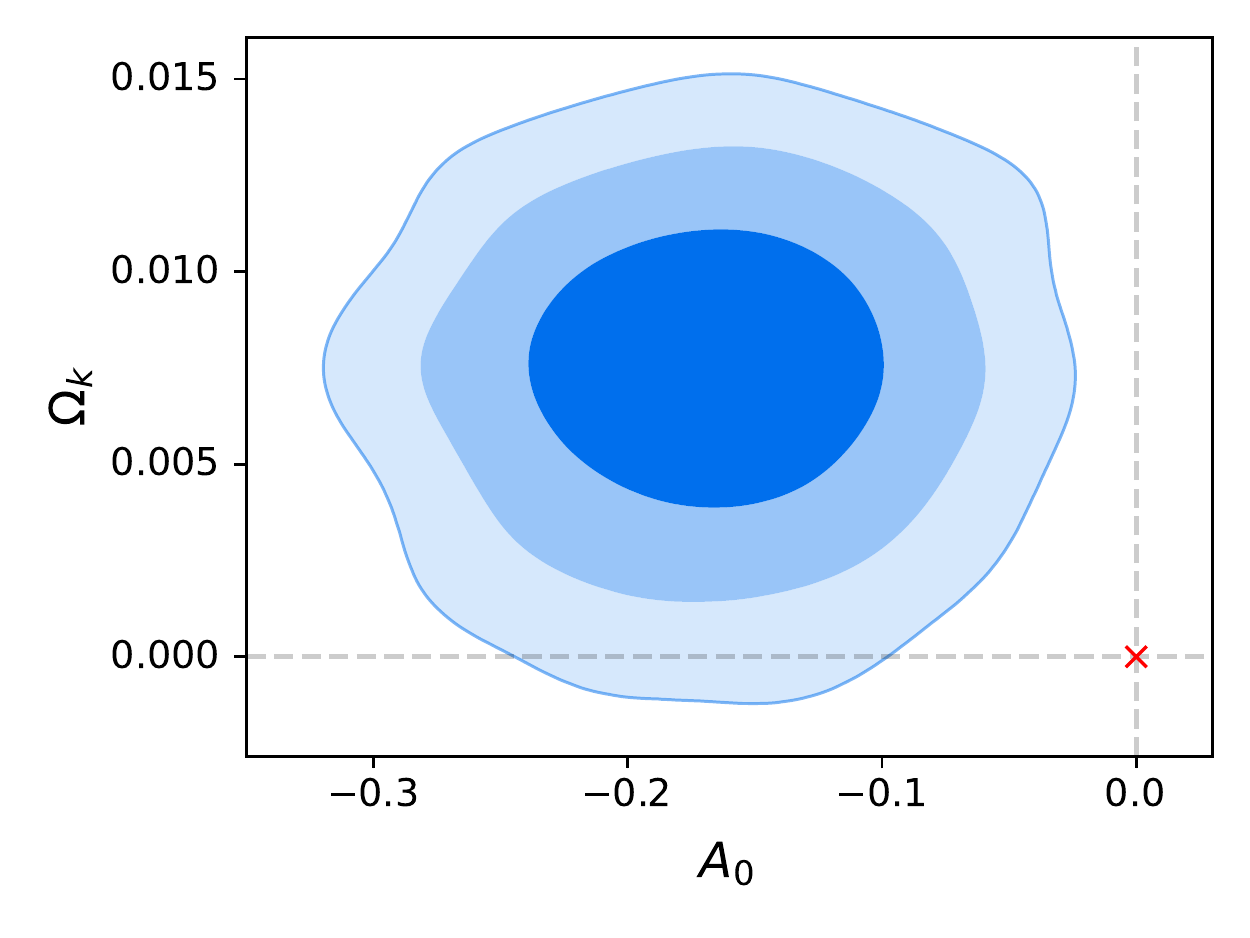}
    \caption{2D posterior probability distributions for parameters $(A_0, \Omega_k)$ in the non-flat $\slcdm$ model. The $\lcdm$ values for these parameters: $(A_0=0, \Omega_k=0)$ is shown as a red cross in the 2D plane. Three contours with confidence intervals of 68, 95 and 99.7 percent respectively are shown. It is clear from the figure that non-zero values of $(A_0, \Omega_k)$ in the non-flat $\slcdm$ model are preferred by the data at more than $3\sigma$ confidence. }
    \label{fig:A0omk}
\end{figure}

In Figure \ref{fig:A0omk}, we show the 2D posterior probability distribution for parameters $(A_0, \omk)$ produced in the non-flat $\slcdm$ fit using the ${\it Planck+SH0ES+Pantheon}$ data. In the figure, the dashed lines show the value of these parameters in the standard $\lcdm$ model i.e. $A_0 =0$ and  $\omk=0$. The $\lcdm$ value in the 2D plane is shown as a red cross at $(A_0=0, \omk=0)$. As can be seen in the figure, the $\lcdm$ value is outside the $3\sigma$ ($99.7 \%$) confidence contour. If we add a $99.95 \%$ confidence contour in Figure \ref{fig:A0omk} which corresponds to about $3.5\sigma$, the $\lcdm$ value (red cross) still lies comfortably outside the contour. However, we start to observe that the contour has a noisier shape indicating that we need a larger sample of MCMC points to make robust inference at $99.95 \%$ confidence level; therefore, we do not plot beyond the $99.7\%$ confidence contour.

\section{Discussion}
We now discuss the improvement in fits in the non-flat $\slcdm$ model compared to the flat $\lcdm$ model. The model fit improvement of the non-flat $\slcdm$ model over $\lcdm$ model for our data combination is substantial: $\Delta \chi^2 = -23.1$. But it is important to check whether the improvement in overall fit of the data occurs at the expense of significantly reducing the quality of fit to a subset of the data.

\begin{table*}[]
    \centering
    \begin{tabular}{|c | c | c | c | c | c |}
    \hline
    {Likelihood} & \thead{non-flat \\ $\slcdm$} & non-flat $\lcdm$ & flat $\lcdm$ & \thead{flat $\lcdm$ \\ (Planck only)} & \thead{flat $\lcdm$ \\(Planck+SH0ES)}  \\
    \hline
    {planck\_2018\_lowl.TT} & 21.43 & 23.67  & 22.10 & 23.26 & 22.40\\
    { planck\_2018\_lowl.EE} & 395.75 & 396.52  & 396.02 & 396.05 & 397.61 \\
    {planck\_2018\_highl\_plik.TTTEEE} & 2338.82 & 2350.32 & 2349.14 & 2344.6 & 2349.74 \\
    { H0.riess2020} & 0.74 & 1.75  & 13.93 & - & 15.71 \\
    {SN.pantheon} & 1036.08 & 1035.33 & 1034.76 & - & - \\
    \hline
    Total $\chi^2$ & 3792.82 & 3807.59 & 3815.95 & - & - \\
    \hline
    $\Delta \chi^2$ (relative to flat $\lcdm$) & -23.1 & -8.4 & 0 & - & - \\
    \hline \hline
    \end{tabular}
    \caption{$\chi^2$ values for best-fit parameters for individual data likelihoods for several models of interest. The data combination used is {\it Planck + SH0ES + Pantheon} except when indicated in the column heading.}
    \label{table:chisq}
\end{table*}

\begin{figure}
	\centering
	\includegraphics[width=\textwidth]{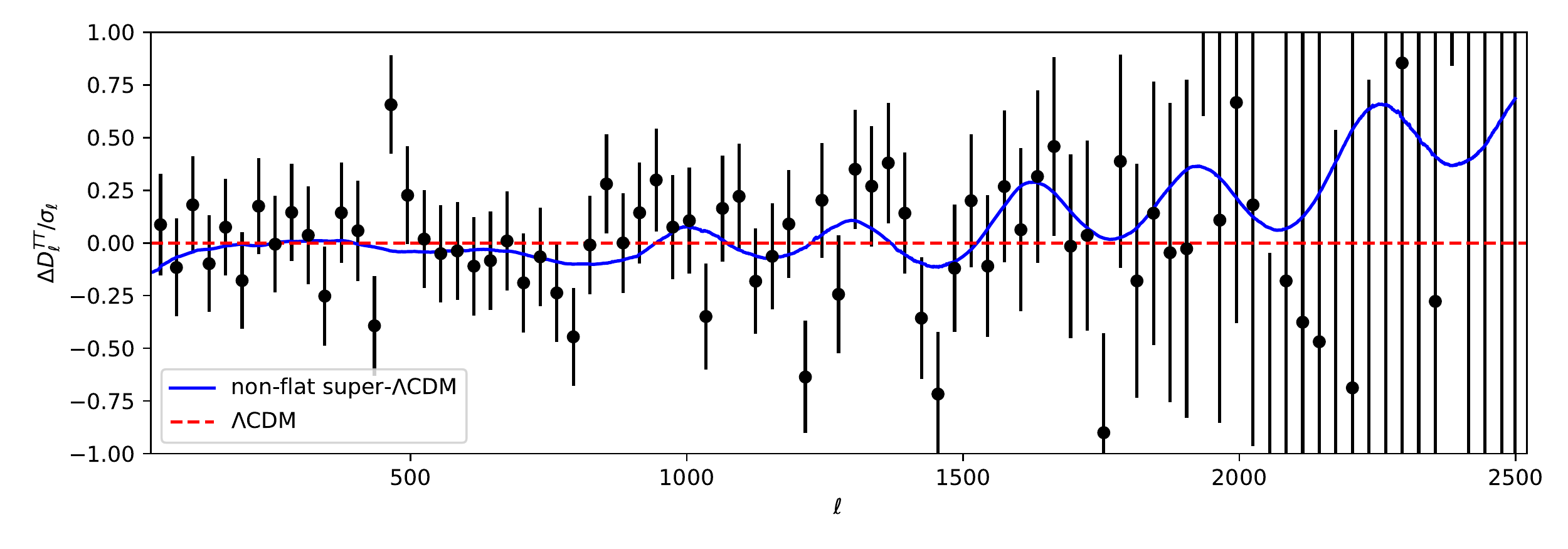}
	\includegraphics[width=\textwidth]{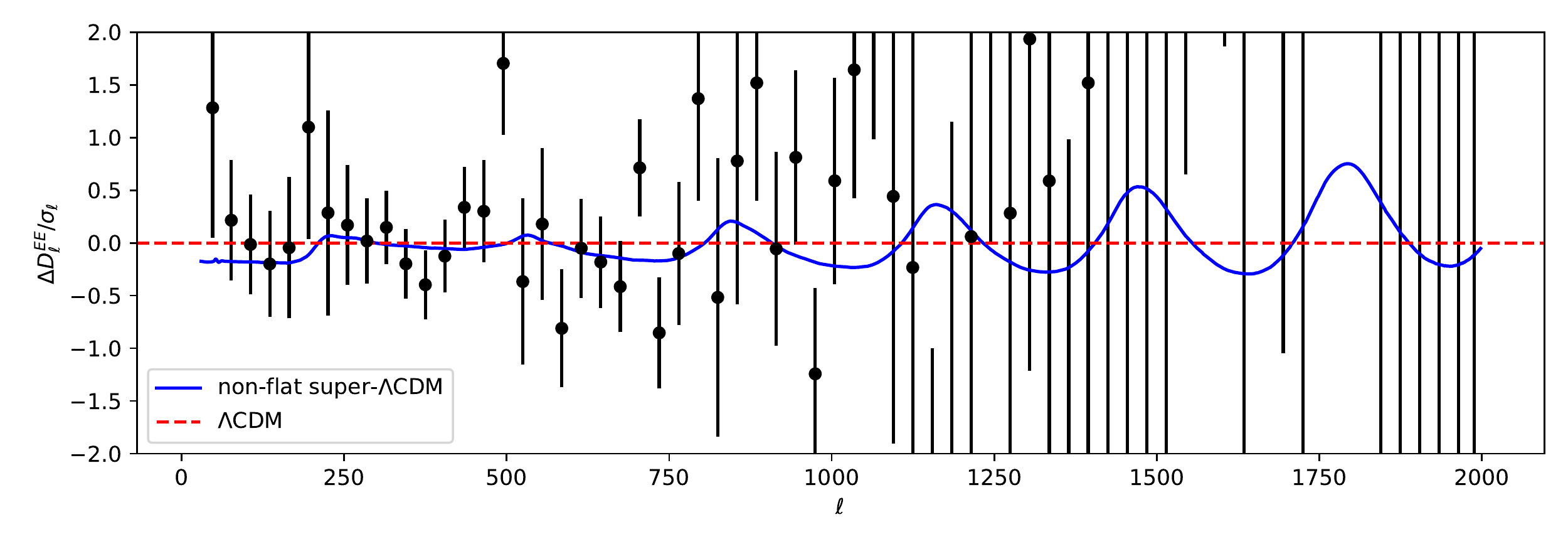}
	\includegraphics[width=\textwidth]{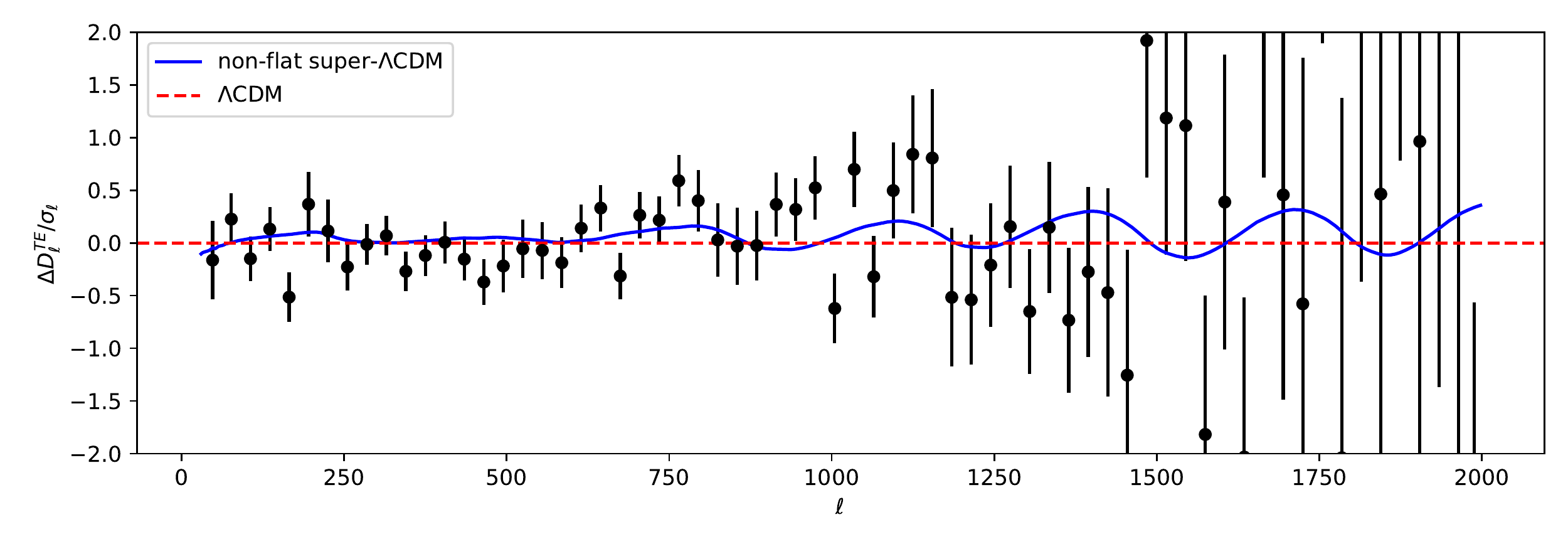}
	\caption{Residuals of Planck $TT, EE,$ and $TE$ angular power spectra with respect to the corresponding best-fit $\lcdm$ theory fit using only {\it Planck} data. The (binned) residuals and error bars plotted above are normalized by cosmic variance. The corresponding best-fit non-flat $\slcdm$ theory residual (blue line) with respect to the best-fit Planck $\lcdm$ theory is also plotted. It can be observed that the non-flat $\slcdm$ model follows the oscillatory features in the temperature power spectrum residuals between multipoles $\ell \approx 800$ and $\ell \approx 1800$. This likely accounts for some of the improvement in CMB $\Delta \chi^2$ in the non-flat $\slcdm$ model.}
	\label{fig:residuals}
\end{figure}

In Table \ref{table:chisq}, we list and compare the $\chi^2$ of fits for different models and data. In the table we can see if any improvement in fits is largely driven by one of the data at the expense of significantly degrading the quality of fit to other data. Comparing the $\chi^2$ of fits for the non-flat $\slcdm$ (second column in Table \ref{table:chisq}) with the $\chi^2$ of fits for the flat $\lcdm$ (fourth column in Table \ref{table:chisq}), it is clear that the non-flat $\slcdm$ model's large improvement in fit of the ${\it Planck+H_0+Pantheon}$ data  does not deteriorate individual {\it Planck} temperature and polarization likelihood fits. The large improvement in $\Delta \chi^2$ is also consistent with the the 2D posterior plot shown in Figure \ref{fig:A0omk}, in which we can observe that the combination of data ${\it Planck + SH0ES + Pantheon}$ prefers a non-zero value of $\omk$ at a statistical significance exceeding $3\sigma$. We note, however, that the preference for non-zero $\omk$ is for the sign opposite to what is found with the ${\it Planck}$-only data in the $\lcdm+\omk$ model in \cite{DiValentino:2019qzk}, which finds a preference for $\omk<0$ whereas our analysis with a different data combination finds a preference for $\omk>0$. 

The mild preference of the {\it Planck} CMB data for negative $\omk$ in the $\lcdm+\omk$ model seems to come from its ability to better fit the apparent larger lensing effect in the temperature data (sometimes parameterized by the phenomenological $A_{\rm lens}$ parameter) in addition to the better fit to the low-$\ell$ temperature data \cite{Planck:2018vyg}. However, those fits were done without adding the SH0ES data. In the fits with {\it Planck+SH0ES+Pantheon} data, we find an improvement in fit in the $\lcdm+\omk$ model over the $\lcdm$ model by $\Delta \chi^2 = -8.4$, but with a preference for $\omk>0$. As can be seen in Table \ref{table:chisq}, most of the improvement in fit is due to better fitting the SH0ES data, while the fit to the {\it Planck} CMB data worsens by $\Delta \chi^2 = 6.6$ (compared to flat $\lcdm$ using {\it Planck}-only data). This is not desirable and points to the fact that the two data sets are discrepant in the $\lcdm+\omk$ model. On the contrary, the non-flat $\slcdm$ model improves fitting of the {\it Planck} CMB data compared to the flat $\lcdm$ model (with {\it Planck}-only data) by $\Delta \chi^2 = -7.9$.

In Figure \ref{fig:residuals}, we plot the residuals of {\it Planck} power spectrum data. The residuals of $D_\ell = \ell(\ell+1) C_\ell/(2 \pi)$ for $TT, EE,$ and $TE$ spectra are calculated with respect to the corresponding best-fit $\lcdm$ Planck 2018 Planck spectra provided by the Planck Colaboration\footnote{\texttt{COM\_PowerSpect\_CMB-base-plikHM\_TTTEEE-lowl-lowE-lensing-minimum-theory\_R3.01.txt}} and plotted after normalizing by the cosmic variance for each multipole. In the figure we also plot the corresponding residuals of the best-fit non-flat $\slcdm$ power spectra; the best-fit model parameters used can be found in Table \ref{table:parameters}. For the $TT$ spectra, we can observe that the non-flat $\slcdm$ model captures some of the oscillatory features of the data residuals ($800\lesssim\ell\lesssim 1800$) leading to an improvement in CMB $\Delta \chi^2$.

\section{Summary}\label{sec:summary}
In this work, we present a concrete example of how the Hubble tension could be a hint of two different extensions to the standard $\lcdm$ model. We find that a non-flat $\slcdm$ model can significantly alleviate the Hubble tension. We find that the model greatly improves the fit to the combination of {\it Planck} temperature plus polarization, the local measurement of the Hubble constant, and the Pantheon supernovae distance measurements. With three extra model parameters, the fit improvement of the $\slcdm$ model with respect to the standard $\lcdm$ for the data combination {\it Planck + SH0ES + Pantheon} is found to be $\Delta \chi^2 = -23.1$. The Hubble constant value preferred is larger compared to the {\it Planck}-only $\lcdm$ derived Hubble constant. We find that the better fit is obtained without degrading the fit to the {\it Planck} likelihood. 

\section*{Acknowledgments}
The computational work necessary for this paper was done using a {\it Google Cloud Research Credits program} with the award GCP19980904. The author would like to thank the anonymous referee for useful comments that has improved the content and presentation of the paper.

\bibliographystyle{elsarticle-num}

\end{document}